# A high throughput (>90%), large compensation range, single-prism femtosecond pulse compressor

Lingjie Kong, Meng Cui*
Janelia Farm Research Campus, Howard Hughes Medical Institute, Ashburn, Virginia 20147, USA
* cuim@janelia.hhmi.org

**Abstract**: We demonstrate a high throughput, large compensation range, single-prism femtosecond pulse compressor, using a single prism and two roof mirrors. The compressor has zero angular dispersion, zero spatial dispersion, zero pulse-front tilt, and unity magnification. The high efficiency is achieved by adopting two roof mirrors as the retroreflectors. We experimentally achieved ~ -14500 $fs^2$ group delay dispersion (GDD) with 30 cm of prism tip-roof mirror prism separation, and ~90.7% system throughput with the current implementation. With better components, the throughput can be even higher.

## 1. Introduction

Femtosecond pulses with high peak-power have found broad applications in micromachining, biomedical imaging, and spectroscopy[1, 2]. However, femtosecond pulses are susceptible to GDD when they propagate through optical elements, resulting in longer pulse duration and lower peak power. Generally, pulse compressors are used to compensate the material dispersion. In multi-photon microscopy, femtosecond pulses with high peak-power are expected to achieve high excitation efficiency[2]. The excitation pulses should be prechirped by pulse compressors before entering the microscope, such that the negative GDD introduced by the compressor cancels the positive GDD of the microscope components.

Components with angle dispersion, such as gratings and prisms, are commonly used to introduce negative GDD. So far, various schemes of femtosecond pulse compressor (with either gratings, prisms, or phase compensation based on SLM or deformable mirrors) have been demonstrated [3-6]. The most common prism compressor is composed of two prisms and a mirror. Although this scheme is much simplified compared to the four-prism pulse compressor, it is still difficult to tune, align and vary its GDD over a wide range. Recently, a single-prism pulse compressor has been demonstrated[1]. The system is compact and easy to align. However, the system throughputis low (~70%) mainly due to the corner cube in the setup.

Here we report a high throughput, large compensation range, single-prism femtosecond pulse compressor. We use roof mirrors (total internal reflection), instead of corner cubes, to ensure high throughput. The system keeps all the advantages of single-prism compressor as in Ref.[1], such as zero angular dispersion, zero spatial dispersion, zero pulse-front tilt, and unity magnification. With ~30 cm of prism tip-roof mirror separation, ~14500 $fs^2$ GDD can be compensated. Current implementation shows ~90.7% system throughput, despite that one silver mirror is used in the system. Replacing the silver mirror with a dielectric mirror can provide a ~3% improvement.

## 2. High throughput, large compensation range, single-prism femtosecond pulse compressor

The proposed compressor is shown in Fig. 1, which is composed of a single prism and two roof mirrors. A laser beam of *p*-polarization enters the dispersive prism at the Brewster angle and becomes spatially dispersed. A roof mirror (90 degree fused silica prism) works as a retroreflector and shifts the beam horizontally. After the second pass through the prism, the beam enters the other roof mirror (90 degree fused silica prism) that shifts the reflected beam vertically (changing the beam height). The reflected beam propagates through the prism twice and is picked off by a D-shape silver mirror.

It should be noted that the roof mirrors adopted in our scheme ensure a higher throughput, compared to that with the corner cube [1]. It is known that corner cubes based on total internal reflection will alter the polarization states of the input beam[7], which decreases the transmission at the prism surface. To keep pulse's polarization state intact, a corner cube with metal coating are used in Ref.[1]. However it is still lossy, as every reflection introduces ~3% loss for pulses at 700~1100 nm. Six reflections will lead to as much as ~17% loss. Here, we use roof mirrors based on total internal reflection as ideal $180^o$ retroreflectors with high reflectivity and zero polarization rotation.

The GDD can be tuned over a broad range by translating the first roof mirror (roof mirror 1 in Fig. 1). Same as in Ref.[1], this scheme is the spatial-temporal distortion free: there is zero angular dispersion, zero spatial dispersion, and zero pulse-front tilt in the output beam, which is experimentally verified with FROG.

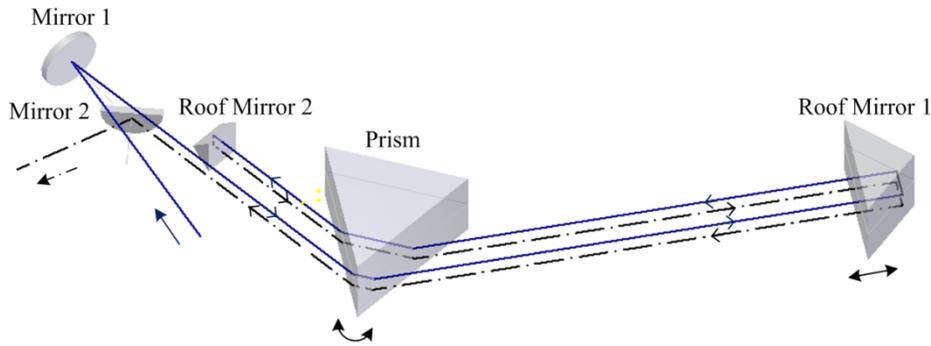

Figure 1 Scheme of the highly efficient single-prism pulse compressor. Beam traces at two different heights are shown with solid line and dash-dot line, respectively.

### 3. Experimental results and discussion

In experiments, we use an isosceles Brewster angle N-SF66 prism (for 930 nm) to disperse light. The fused silica roof mirrors' hypotenuses are anti-reflection coated. Mirror 1 and 2 in Fig.1 are dielectric mirror and D-shape silver mirror, respectively. The laser pulses at 930 nm are from a Ti:Sapphire oscillator (Chameleon, Coherent), and the pulses are measured with GRENOUILLE[8] (Swamp Optics) before and after the pulse compressor. Figure 2 shows the measured GDD at different prism tip-roof mirror separations. Linear fitting shows that the slope of GDD versus the prism tip-roof mirror separation is -84 $fs^2$/mm (95% confidence bounds), agreeing well with the calculated result -82 $fs^2$/mm (Lab2 [9]). With ~30 cm of prism tip-roof mirror separation, ~14500 $fs^2$ GDD can be compensated. This is sufficient to compensate the dispersion in even highly complicated microscopes[1, 10].

The system throughput is 90.7±0.7% for pulses at 930 nm. Replacing the D-shape silver mirror with a HR coated dielectric mirror would provide ~3% improvement.

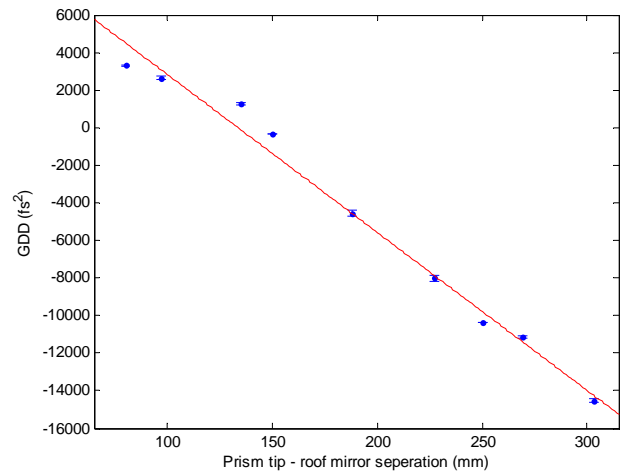

Figure 2 GDD at different prism tip-roof mirror separations. Red line: linear fitting.

## 4. Conclusion

We proposed and demonstrated a high throughput, large compensation range, single-prism femtosecond pulse compressor. It is of zero angular dispersion, zero spatial dispersion, zero pulse-front tilt, and unity magnification. Fused silica roof mirrors are adopted to ensure high throughput of the system. With current design, ~14500 $fs^2$ GDD can be compensated with ~30 cm of prism tip-roof mirror separation and the system throughput reaches ~90.7%. Replacing the silver mirror with a dielectric mirror will provide additional ~3% improvement.

## Acknowledgement

This research is supported by Howard Hughes Medical Institute.